\documentclass[published]{nst}

\usepackage{subfigure,dcolumn}
\usepackage{siunitx}
\usepackage{amssymb}
\usepackage[commandnameprefix=always, markup=default,authormarkup=none, final]{changes}
\usepackage{makecell}
\usepackage[T2A,T1]{fontenc}
\usepackage[russian,english]{babel}

\definechangesauthor[name={Ruiyang Zhang}, color=orange]{rzhang}

\usepackage{listings}
\begin{document}

\title{A Low Background Beta Detection System using a Time Projection Chamber}\thanks{This work was supported by the National Natural Science Foundation of China~(No. 12125505), and the State Key Laboratory of Particle Detection and Electronics~(No. SKLPDE-ZZ-202401)}

\author{Rui-Yang Zhang}
\affiliation{State Key Laboratory of Particle Detection and Electronics,\\
University of Science and Technology of China, Hefei 230026, China}
\affiliation{Department of Modern Physics, University of Science and Technology of China, Hefei 230026, China}
\author{Zhi-Yong Zhang}
\email[Corresponding author, ]{Zhi-Yong Zhang\\
University of Science and Technology of China, Hefei 230026, China \\
+86 18949895439 \\
zhzhy@ustc.edu.cn}
\affiliation{State Key Laboratory of Particle Detection and Electronics,\\
University of Science and Technology of China, Hefei 230026, China}
\affiliation{Department of Modern Physics, University of Science and Technology of China, Hefei 230026, China}
\author{Zeng-Xuan Huang}
\affiliation{State Key Laboratory of Particle Detection and Electronics,\\
University of Science and Technology of China, Hefei 230026, China}
\affiliation{Department of Modern Physics, University of Science and Technology of China, Hefei 230026, China}
\author{Yong Zhou}
\affiliation{State Key Laboratory of Particle Detection and Electronics,\\
University of Science and Technology of China, Hefei 230026, China}
\affiliation{Department of Modern Physics, University of Science and Technology of China, Hefei 230026, China}
\author{Jian-Bei Liu}
\affiliation{State Key Laboratory of Particle Detection and Electronics,\\
University of Science and Technology of China, Hefei 230026, China}
\affiliation{Department of Modern Physics, University of Science and Technology of China, Hefei 230026, China}
\author{Song-Song Tang}
\affiliation{State Key Laboratory of Particle Detection and Electronics,\\
University of Science and Technology of China, Hefei 230026, China}
\affiliation{School of Nuclear Science and Technology, University of Science and Technology of China, Hefei 230026, China}
\author{Yuan-Fei Cheng}
\affiliation{State Key Laboratory of Particle Detection and Electronics,\\
University of Science and Technology of China, Hefei 230026, China}
\affiliation{Department of Modern Physics, University of Science and Technology of China, Hefei 230026, China}
\author{Chang-Qing Feng}
\affiliation{State Key Laboratory of Particle Detection and Electronics,\\
University of Science and Technology of China, Hefei 230026, China}
\affiliation{Department of Modern Physics, University of Science and Technology of China, Hefei 230026, China}
\author{Ming Shao}
\affiliation{State Key Laboratory of Particle Detection and Electronics,\\
University of Science and Technology of China, Hefei 230026, China}
\affiliation{Department of Modern Physics, University of Science and Technology of China, Hefei 230026, China}
\author{Yi Zhou}
\affiliation{State Key Laboratory of Particle Detection and Electronics,\\
University of Science and Technology of China, Hefei 230026, China}
\affiliation{Department of Modern Physics, University of Science and Technology of China, Hefei 230026, China}

\begin{abstract}
In this paper, we present a Time Projection Chamber (TPC) system for low-background beta radiation measurements. The system consists of a TPC with two-dimensional-strip readout Micromegas and an anti-coincidence detector with readout pads for cosmic ray veto. The detector system utilize an AGET-based waveform sampling system for data acquisition.

The beta detection capability of the system was verified through experimental test using $^{90}$Sr beta source. Additionally, a dedicated simulation program based on Geant4 was developed to model the entire detection process, including responses to both the beta source and background radiation. Simulation results were compared with experimental data for both beta and background samples, showing good agreements.

The simulation samples were utilized to optimize and train classification models for beta and background discrimination. By applying the selected model into test data, the system achieved a background rate of \SI{0.49}{cpm/cm^2} while retaining more than 55\% of $^{90}$Sr beta signals within a \SI{7}{cm} diameter detection region. Further analysis revealed that approximately 70\% of the background originates from environmental gamma radiation, while the remaining contribution mainly comes from intrinsic radioactivity of detector materials, particularly the FR-4 based field cage and readout plane. Based on the knowledge gained from the experiments and simulations, an optimization of the TPC system has been proposed, with simulation predicting a potential reduction of the background rate to \SI{0.0012}{cpm/cm^2}.
\end{abstract}

\keywords{Gaseous detector, Time Projection Chamber, Micromegas, Low background beta detection, Detector modelling and simulations, Data processing methods}

\maketitle

\section{Introduction}

Since the discovery of alpha and beta radiation by Rutherford in 1899~\cite{RN27}, their properties have been extensively studied and widely utilized in nuclear physics experiments. Alpha and beta radiation are commonly found in the environment and are generally originated from natural and human produced radionuclides~\cite{shahbazi2013review}. In specific situations, it is crucial to monitor radiation levels to ensure they remain within safety standards~\cite{reactors}. For instance, measurements of $^{90}$Sr radioactivity in environmental samples near the Fukushima Daiichi Nuclear Power Plant in 2013 provided insights into the nuclide’s mitigation following the nuclear accident~\cite{RN61}. Similarly, radioactivity of alpha and beta emitting sources were monitored at Qinshan Nuclear Power Plant for a long period of time~\cite{qinshan, qinshan2}. In rare-event search experiments, alpha and beta radiation from uranium and thorium decay chains can significantly contribute to background signals~\cite{formaggio2004backgrounds}. Evaluating and suppressing these backgrounds is critical for achieving the desired sensitivity, as demonstrated in experiments like \chreplaced[id=rzhang]{XENON~\cite{aprile2011study}, NEXT~\cite{novella2019radiogenic} and N$\nu$DEx~\cite{nvdex}}{XENON~\cite{aprile2011study} and NEXT~\cite{novella2019radiogenic}}. Beyond nuclear and high-energy physics, alpha and beta radiation detection is also essential for monitoring pollution in drinking water and food~\cite{water1, water2, food_review, food1, food2}. In all these scenarios, accurately determining radiation levels is vital.

However, radiation signals from samples are often accompanied by natural background radiation, necessitating the development of low-background detection techniques for precise measurements. The primary sources of background include environmental gamma rays, cosmic rays, and radioactive contamination in construction materials, with their contributions varying depending on the detection techniques and the type of radiation being measured~\cite{background_source, low_background_techniques}. While several solutions exist in pursuing low-background radiation measurement~\cite{jobbagy2010current}, including commercially available detectors based on proportional counters or scintillators~\cite{RN45, MPC9604, scint, scint2}, these techniques rely primarily on counting rates and pulse height information to distinguish between different radiation types. This approach imposes limitations, particularly when detecting beta radiation, as its energy loss is similar to that of environmental background radiation. Consequently, these detectors lack the ability to actively separate beta signals from background noise, often requiring thick lead shielding layers (typically \SI{10}{\centi\meter}) to achieve low background rates. Such shielding materials restrict the detection volume and compromise system portability.

To address these limitations, we propose using a Time Projection Chamber (TPC) as the foundational concept for low-background alpha and beta radiation detection~\cite{RN37}. TPC offers the capability to simultaneously measure the 3D track and energy deposition of charged particles within the detector. By leveraging the detailed information provided by the incoming particles, it becomes possible to effectively discriminate target particles of interest from background signals. Furthermore, compared to traditional proportional counter-based techniques, TPC provides the additional advantage of event localization, enabling the determination of the spatial distribution of radiation within the sample.

In our previous work~\cite{RN37}, a Micromegas-based time projection chamber (TPC) prototype was developed for low-background alpha detection. The results demonstrated an alpha background rate of less than \SI{1.6e-3}{cpm} at a 95\% confidence level without shielding, while achieving a $^{241}$Am acceptance rate of 96\%. \chadded[id=rzhang]{The excellent performance in alpha particle detection highlights the potential of TPC technology for low-background radiation detection~\cite{screener3D,alphaTPC}.} Building on this success, we have developed a new TPC system aimed at low-background beta radiation detection. Unlike alpha particles, beta particles have significantly weaker ionization power, resulting in longer track lengths~\cite{siegbahn2012alpha}. Furthermore, beta detection is subject to additional background sources, such as cosmic rays and environmental gamma radiation. To address these challenges, we optimized the TPC system design to improve its sensitivity and discrimination capabilities. A dedicated simulation program was established to model the detector response, providing insights into its performance and the separation of beta signals from background events.

In this paper, we present the design, fabrication, simulation, and test results of the TPC prototype for beta radiation detection in Section~\ref{sec2}–\ref{sec5}. The beta detection background origins were investigated in Section~\ref{sec6} using both experimental data and simulation results. Finally, further optimizations of the detector are discussed in Section~\ref{sec7}, with simulations predicting an anticipated reduction in the background rate.



\section{Detector structure}
\label{sec2}
\begin{figure}[htbp]
\centering
\includegraphics[width=\hsize]{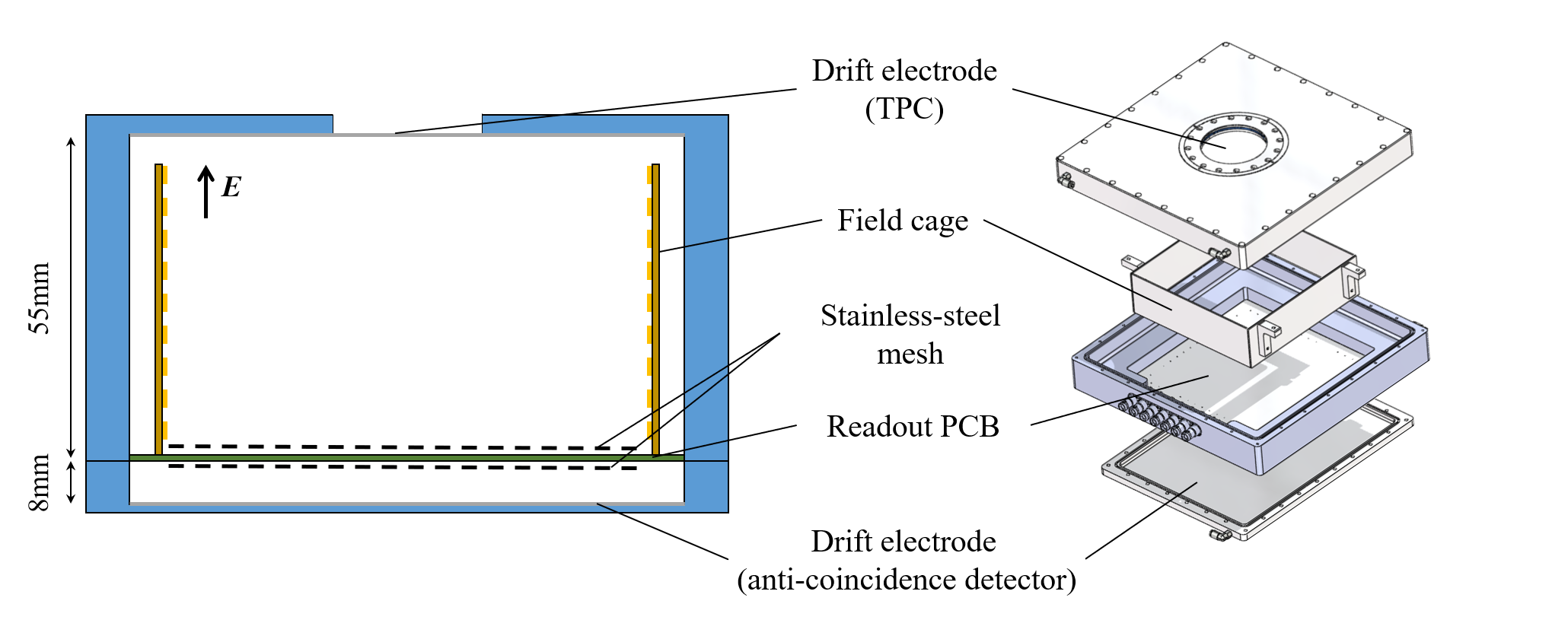}
\caption{Schematic lay-out (left) and the mechanical drawing (right) of the TPC system. \label{fig1}}
\end{figure}

The new TPC prototype builds upon experience gained from the previous version and is optimized to achieve a larger detection area while effectively suppressing background in beta radiation detection. The system consists of two sub-detectors assembled back-to-back: a TPC detector and an anti-coincidence detector, as shown in Figure~\ref{fig1}. The TPC detector serves as the primary detection element for beta particles, while the anti-coincidence detector is employed to veto cosmic ray background. The drift lengths of the TPC detector and the anti-coincidence detector are \SI{55}{\milli\meter} and \SI{8}{\milli\meter}, respectively. For clarity, the term ``TPC'' will refer to the TPC detector alone, while ``TPC system'' will refer to the entire combined detection system hereafter.

\begin{figure}[htbp]
\centering
\includegraphics[width=\hsize]{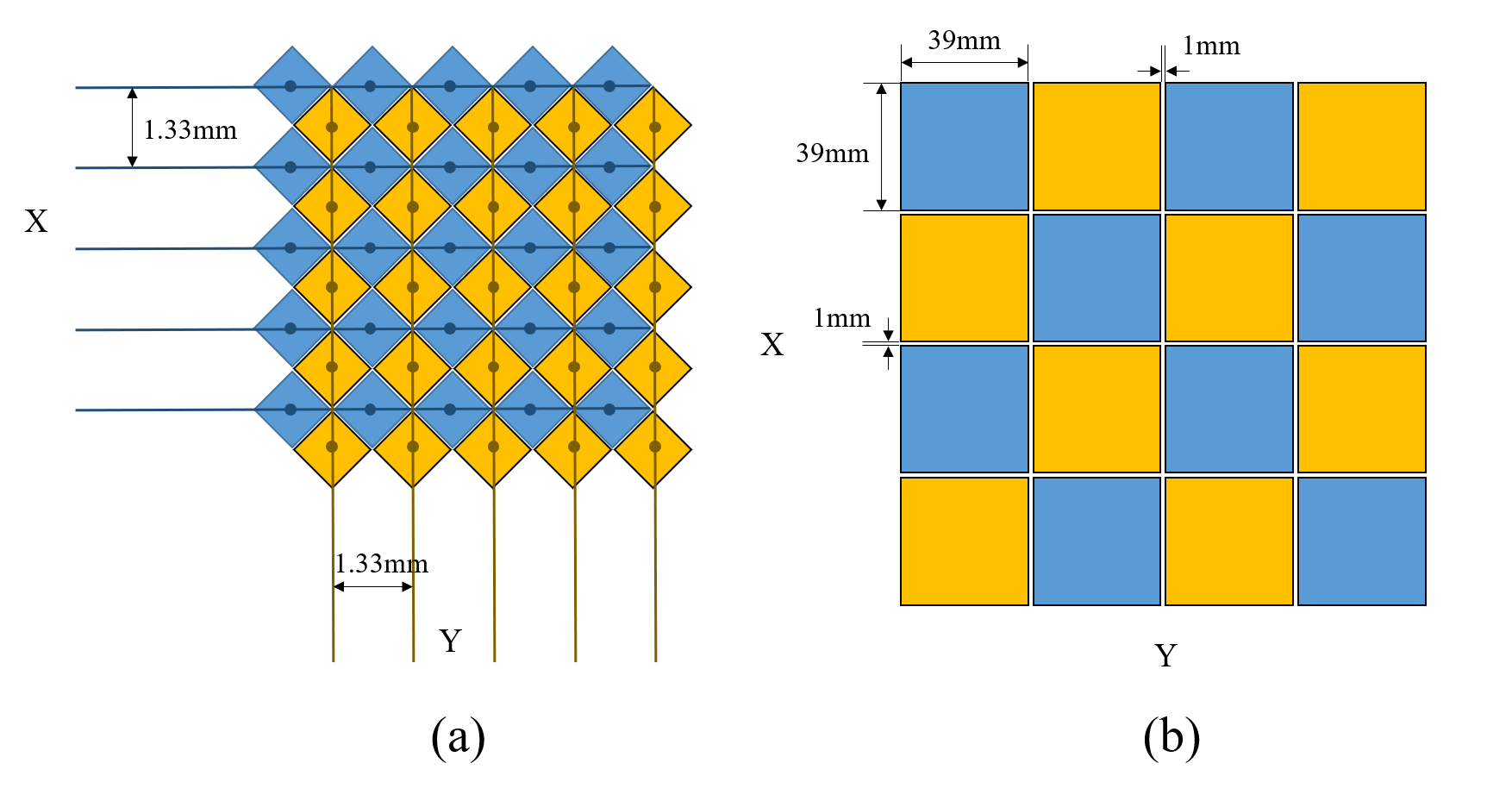}
\caption{Arrangement of the TPC's readout strips (a) and the anti-coincidence detector's readout pads (b). \label{fig2}}
\end{figure}

Thermal-bonded Micromegas~\cite{RN31, RN32} is used both as the readout plane of the TPC and as the anti-coincidence detector. The core component of Micromegas comprises a stainless-steel mesh bonded to a printed circuit board (PCB) with thermal bonding adhesive, forming a \ensuremath{\SI{100}{\micro\meter}} amplification gap. The readout electrodes of the two detectors are situated on opposite sides of the PCB. The PCB is a 10-layer board with a thickness of \SI{2.3}{\milli\meter}, sufficient to block beta particles with energies below \SI{5}{MeV} from reaching the anti-coincidence detector. To provide discharge protection, a germanium layer with a sheet resistivity of approximately \SI{100}{\mega\ohm}/$\square$ is coated on top of the readout electrodes~\cite{feng2022novel}. The readout electrode structures differ between the two sub-detectors. For the TPC detector, the signals are read out through two sets of orthogonally arranged strips. These two-dimensional strips are positioned on the same layer of the PCB and collectively cover the entire detection plane (see Figure~\ref{fig2} (a)).
Each strip in the TPC readout has a width of \SI{1.33}{\milli\meter}, with 120 strips per dimension, resulting in a total sensitive area of $\SI{160}{\milli\meter} \times \SI{160}{\milli\meter}$. This design ensured that the weighting field experienced by the two-dimensional strips is identical, producing uniform responses across both dimensions. The spatial resolution of the TPC was later measured using a collimated $^{241}$Am source, yielding a consistent result of approximately \SI{0.6}{\milli\meter} in both the X and Y directions. In contrast, the anti-coincidence detector, primarily used to veto cosmic ray events, adopts a coarser segmentation to minimize the number of required electronic channels. Its $\SI{160}{\milli\meter} \times \SI{160}{\milli\meter}$ sensitive area is divided into a $4 \times 4$ grid of square pads, with each pad read out independently. Each pad is $\SI{39}{\milli\meter} \times \SI{39}{\milli\meter}$ in size, with a \SI{1}{\milli\meter} gap between adjacent pads, shown in Figure~\ref{fig2} (b). The TPC features a circular incident window with a \SI{7}{\centi\meter} diameter. This window is constructed using a \SI{2.7}{\micro\meter} thick PET film coated with aluminum, designed to minimize energy loss as particles traverse it. The film is electrically grounded by connecting it to the detector’s aluminum shell, where it also serves as part of the drift electrode plane. 

\begin{figure}[htbp]
\centering
\includegraphics[width=\hsize]{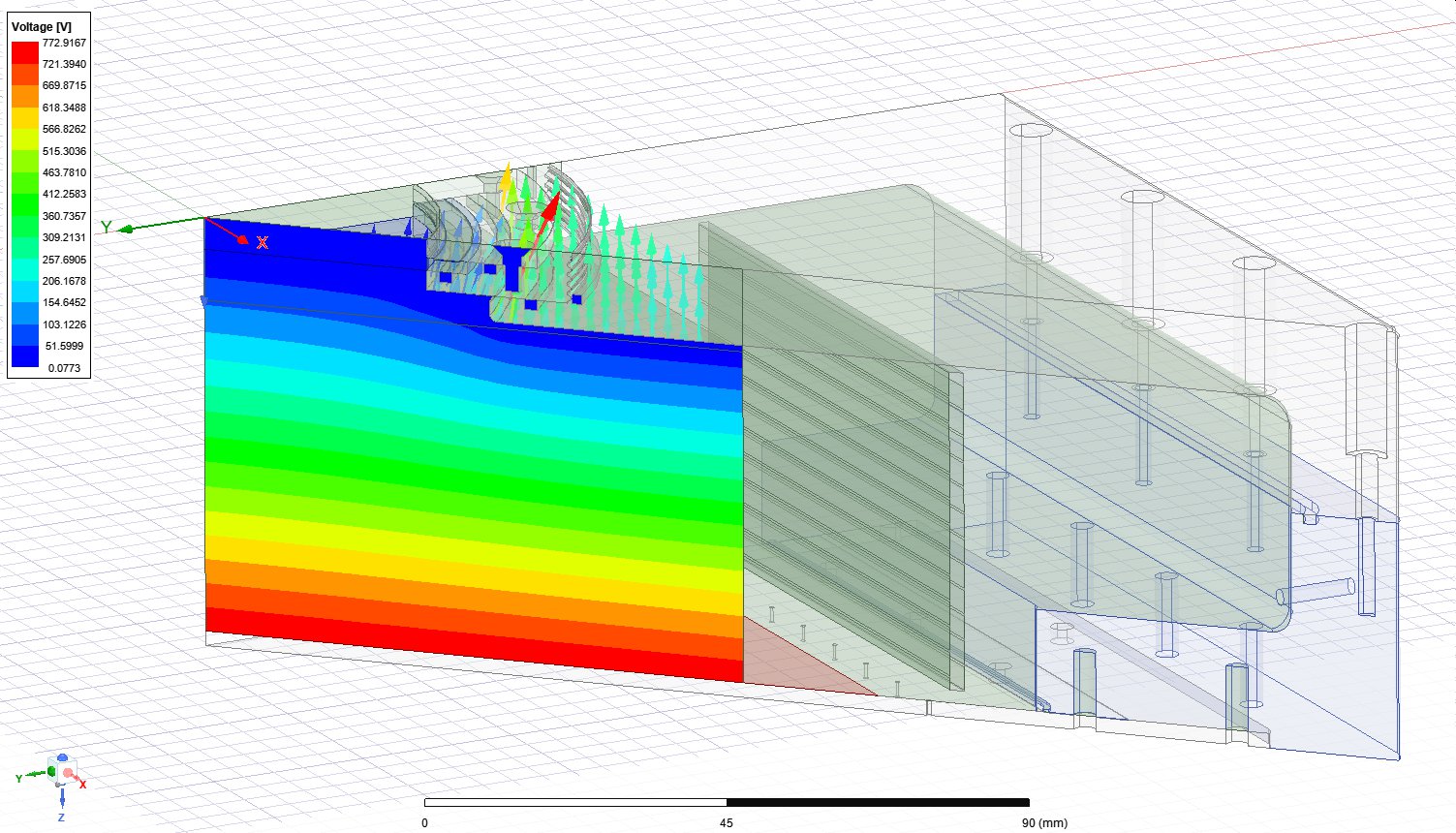}
\caption{Simulated voltage distribution within the central \SI{120}{mm}$\times$\SI{120}{mm} region of the TPC. The voltages correspond to typical settings used during beta radiation detection. Due to the symmetry of the design, only one-eighth of the TPC is modeled. \label{fig3}}
\end{figure}

For the TPC, the design of the field cage is critical to ensuring a uniform electric field within the sensitive volume. Non-uniformities in the electric field can cause abrupt changes in electron drift velocity, affecting track reconstruction accuracy and complicating event discrimination. While this issue is less significant for alpha detection — since alpha particle tracks are short and primarily confined to the central region where field distortion is minimal — it becomes more prominent for beta detection. Beta particles, even with energies of a few tens of kilo-electronvolts, have a higher probability of traversing the entire TPC, making field uniformity essential. To address this, a staggered double-layer strip arrangement was adopted for the field cage~\cite{ILC}. The field cage consists of four PCBs that enclose the sensitive volume. On both the inner and outer surfaces of each PCB, \SI{3.6}{\milli\meter} wide copper strips are arranged with a pitch of \SI{4}{\milli\meter}, with the inner and outer strips displaced by half a pitch. Adjacent strips are connected via resistors to create a uniform voltage gradient when voltage is applied to the bottom strip. The double-layer design provided a key benefit that it mitigates edge effects by shielding external ground voltages. An electromagnetic simulation using ANSYS Maxwell confirmed that, after optimization, the voltage distribution is smooth and uniform across the entire sensitive volume (see Figure~\ref{fig3}).

\section{Experimental setup}
\label{sec3}

\begin{figure}[htbp]
\centering
\includegraphics[width=\hsize]{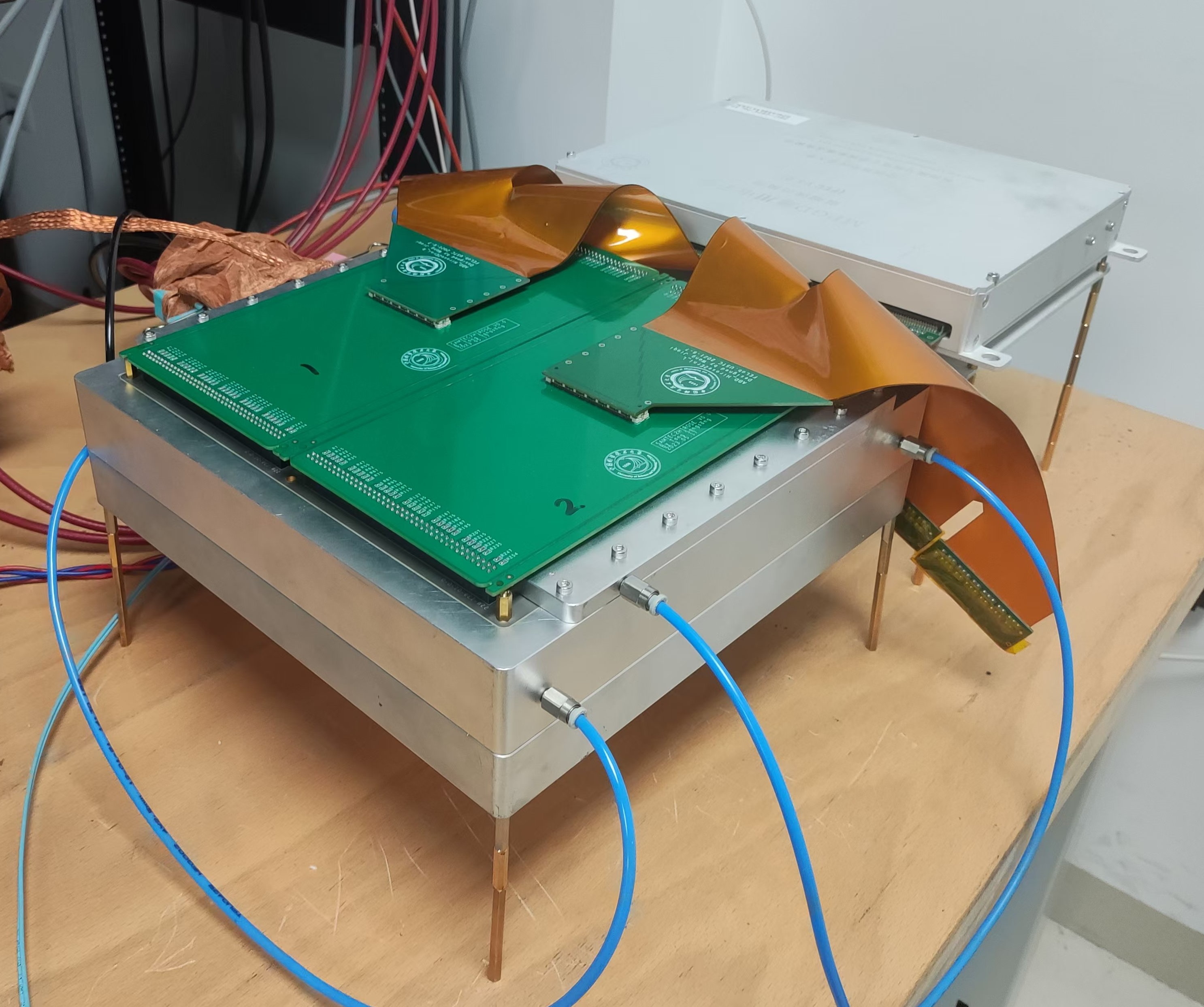}
\caption{A photograph of the TPC system during beta background test. \label{fig4}}
\end{figure}

The TPC system was filled with a gas mixture of 96.5\% argon and 3.5\% isobutane during all tests. The drift cathode was mechanically connected to the detector shell, which automatically set it to ground. The amplification voltage was fixed to \SI{390}{V} after optimization for a desired avalanche gain. The stainless-steel mesh and resistive anode layer were biased at specific positive voltages to maintain an appropriate drift field and a high electron collection efficiency. The anode strips (or pads) of the TPC system were connected to an electronic system for data acquisition and processing. The electronic system used in this experiment is an AGET chip~\cite{anvar2011aget} based multi-channel waveform sampling system developed in USTC~\cite{zhu2019development}. Signals from the anode channels of both the TPC and the anti-coincidence detector were first sent to a front-end card (FEC), where they were amplified, shaped, sampled, and digitized. The shaping time of the FEC was set to \SI{500}{\nano\second}. For each channel, the signal waveform was digitized at a sampling rate of \SI{25}{\mega\hertz} with 512 sampling points, resulting in a time window of \SI{20}{\micro\second}. This time window was chosen to cover the maximum drift time of an ionization track generated by beta particles in the TPC, which is approximately \SI{2}{\micro\second}. The digitized waveform data were then transmitted to a data collection module (DCM), which aggregated information from the FEC and organized signals from different channels into events. Finally, these event packets were sent to a personal computer for further analysis. A photograph of the TPC system and its associated electronics is shown in Figure~\ref{fig4}.

In this study, a $^{90}$Sr surface beta source with a diameter of \SI{1}{\centi\meter} was used to generate reference beta radiation events. The source was placed at the center of the incident window, \SI{5}{\centi\meter} away from the window during the tests. The decay product of $^{90}$Sr is $^{90}$Y, which itself undergoes beta decay. As a result, the recorded events included beta particles from both $^{90}$Sr and $^{90}$Y, with decay energies of \SI{0.546}{\mega\electronvolt} and \SI{2.28}{\mega\electronvolt}, respectively~\cite{nudat3}. In addition to the beta source data, background data were also collected under the same test conditions (voltages, working gas) but without the beta source, in order to assess the background performance of the detector system.

As mentioned earlier, the test data were stored as event packages. Each event package contains the numbers of all hit channels' signal waveforms, from which we can reconstruct the energy deposition and particle track. The amplitude of each waveform is proportional to the charge collected by the channel, and thus also proportional to the energy deposited in the region corresponding to that detector channel. Therefore, the sum of the amplitudes of all channels represents the total energy deposition in the TPC for a given particle. The position of an ionization can be determined by the hit channel's position and the signal's arrival time. Since the TPC uses a two-dimensional strip readout structure, it allows for 2D track reconstruction in both the XZ and YZ planes. 

\begin{figure}[htbp]
\centering
\includegraphics[width=\hsize]{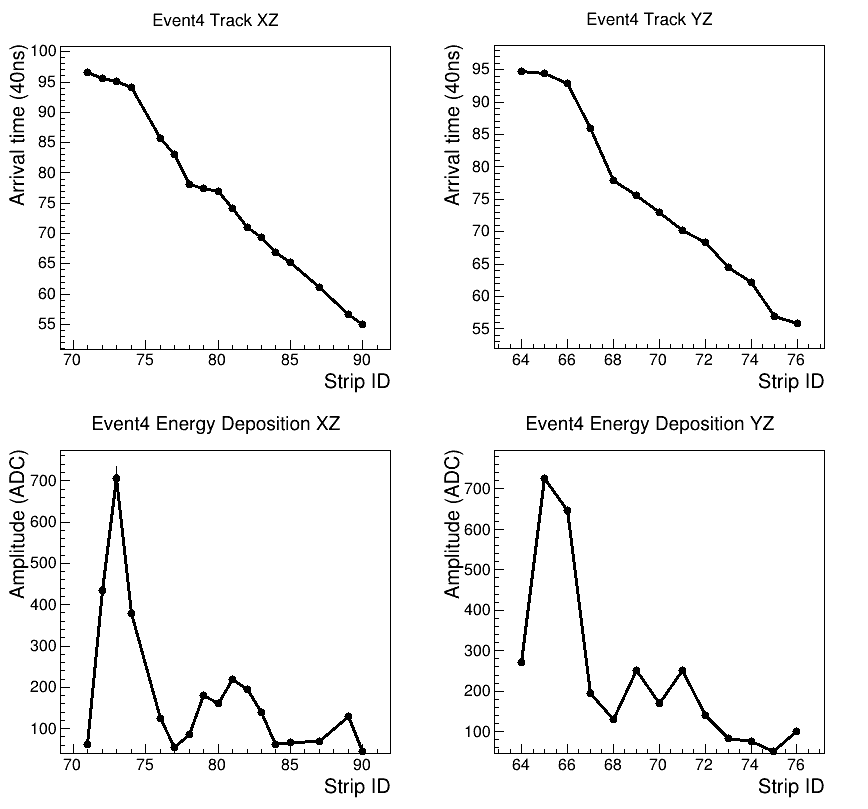}
\caption{A beta event detected by the TPC. Top row: 2D projected tracks on XZ and YZ plane. The Z dimension of the track is denoted by the signal arrival time on each strip. Bottom row: waveform amplitudes on X and Y strips.\label{fig5}}
\end{figure}

\begin{figure}[htbp]
\centering
\includegraphics[width=\hsize]{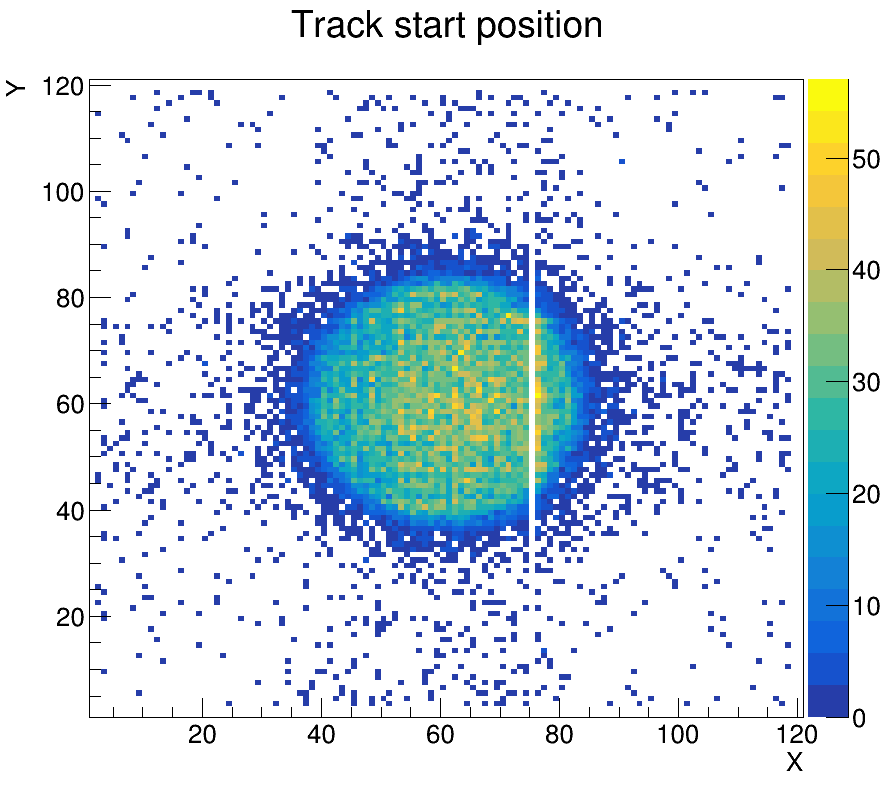}
\caption{Start position map of $^{90}$Sr beta test data.\label{fig6}}
\end{figure}

Figure~\ref{fig5} (top row) shows a typical reconstructed beta particle track projected on XZ and YZ plane, where the particle's scattering due to multiple interactions is illustrated. The Z-dimensional drift length is determined by the arrival time, measured using the 20\% constant fraction timing method on the corresponding strip. The amplitude on each channel, representing the energy deposition, is also shown for reference.
The start position map of the beta particles is shown in Figure~\ref{fig6}. The start position is defined by combining the X and Y strip numbers with the longest drift time for each event, which corresponds to the position where the particle first enters the TPC volume. The reconstructed start positions for beta particles are primarily concentrated in a central circular area, the size of which is consistent with that of the detector's entrance window. This is understandable, as beta particles are easily blocked by the detector frame outside the entrance region. A missing channel in the X dimension is attributed to it being connected to a noisy channel on the AGET chip, so that the output signal from this channel was dropped. These test results indicate that the TPC system is well-capable of detecting beta particles and recording their 3D trajectories.

\section{Detector simulation}
\label{sec4}

As discussed in the previous section, we have obtained test data for the $^{90}$Sr beta source and background events. To further distinguish beta source events from background events, it is essential to gain a deeper understanding of the different behaviors exhibited by beta and background events. For this purpose, a Monte Carlo simulation based on Geant4 software~\cite{geant4} was conducted to model the signal generation process for both the radioactive beta source and background particles.

The detector geometry used in the simulation was constructed according to the TPC system's mechanical design, with some simplifications in the detailed structures. A central region of \SI{160}{\milli\meter}$\times$\SI{160}{\milli\meter} was defined as the sensitive area. The drift gaps for the TPC and the anti-coincidence detector were \SI{55}{\milli\meter} and \SI{8}{\milli\meter}, respectively. A filter was applied to select electrons generated by physical processes other than ionization, ensuring that only primary electron tracks were counted. These included beta particles from beta decay as well as electrons produced by photon interactions with materials.

The aforementioned part of the simulation can be naturally implemented in Geant4. However, the simulation results could not be directly compared with our test data, as the simulation did not account for the detector and electronics response. To bridge this gap, an additional digitization process was incorporated into the simulation program, which converted the Monte Carlo truth information into waveforms on the electronics channels.

For a given simulated electron track, ionized electrons along the track were grouped and drifted towards the anode plane under the electric field. The drift parameters for our working gas mixture (\SI{96.5}{\percent} Ar + \SI{3.5}{\percent} $\rm iC_4H_{10}$) were derived from Garfield++ simulations~\cite{garfield}. Under our test conditions for beta particles and their background, the drift field was set to approximately \SI{120}{\volt\per\centi\meter}, where the electron drift velocity is \SI{3.72}{\centi\meter\per\second}, and the longitudinal and transverse diffusion coefficients are \SI{0.037}{\centi\meter^{1/2}} and \SI{0.055}{\centi\meter^{1/2}}, respectively. The arrival position and time of each drifted electron cluster were calculated assuming a constant drift velocity, with a random Gaussian-distributed displacement both horizontally and vertically, based on the diffusion coefficients.

The drifted electrons induced charge signals on the anode plane following the avalanche process in the Micromegas, which was represented by a constant gain factor. The anode plane of the TPC was divided into X and Y strips, as shown in Figure~\ref{fig2}. To account for the effect of signal spreading on the resistive layer, a charge cluster not only induced a signal on its hit strip, but also on two adjacent strips in both the X and Y dimensions. The total drifted charge was shared between these five strips, with the charge distribution on each strip inversely proportional to its distance from the hit position. The collected charges on each strip were then convoluted with a given transfer function, calibrated for the electronics, to obtain the analog output signal. Finally, this signal was sampled at a rate of \SI{25}{\mega\hertz}, producing the actual output signal waveform with 512 sampling points. This digitization process generated simulation data with the same format as the test data, enabling a direct comparison between the simulation results and test data, and allowing further optimization of the detector design based on the simulation outcomes.

\begin{figure}[htbp]
\centering
\includegraphics[width=\hsize]{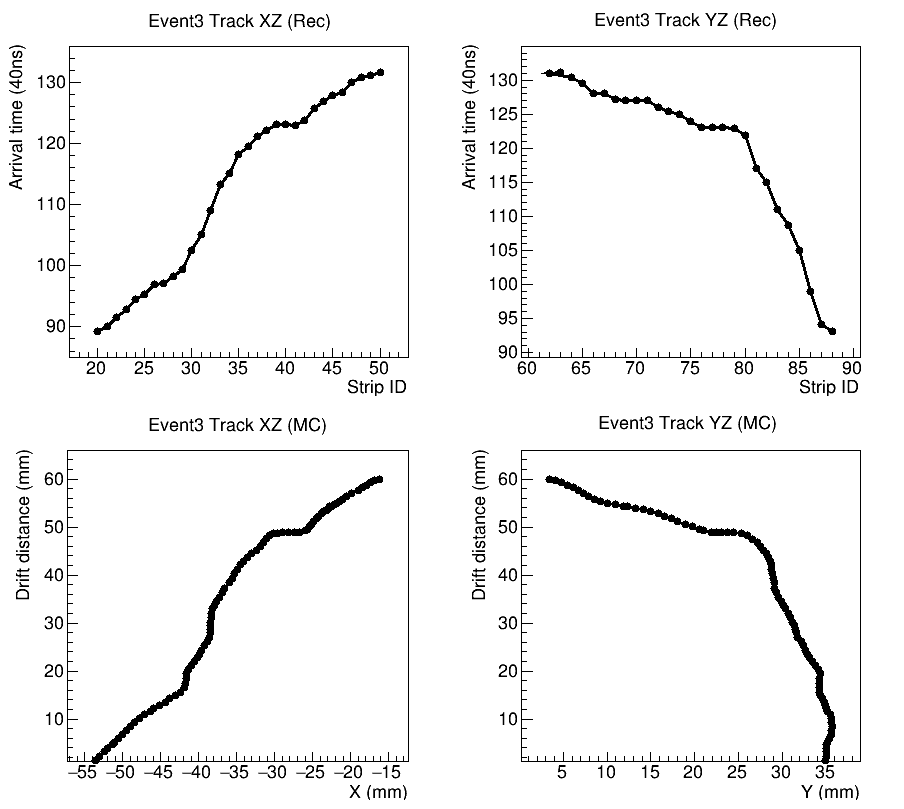}
\caption{Comparison between reconstructed track (top) and MC truth track (bottom). \label{fig7}}
\end{figure}

Figure~\ref{fig7} shows a comparison between the MC truth and the reconstructed track after digitization and data processing. There is almost no difference in the track shape, indicating that the digitization process did not significantly affect the TPC’s tracking performance. Additionally, since the digitization reflects the actual detector response, this result further implies that the designed detector readout segmentation is appropriate and that the multi-scattering effect caused no significant deterioration in track reconstruction.

\begin{figure}[htbp]
\centering
\includegraphics[width=\hsize]{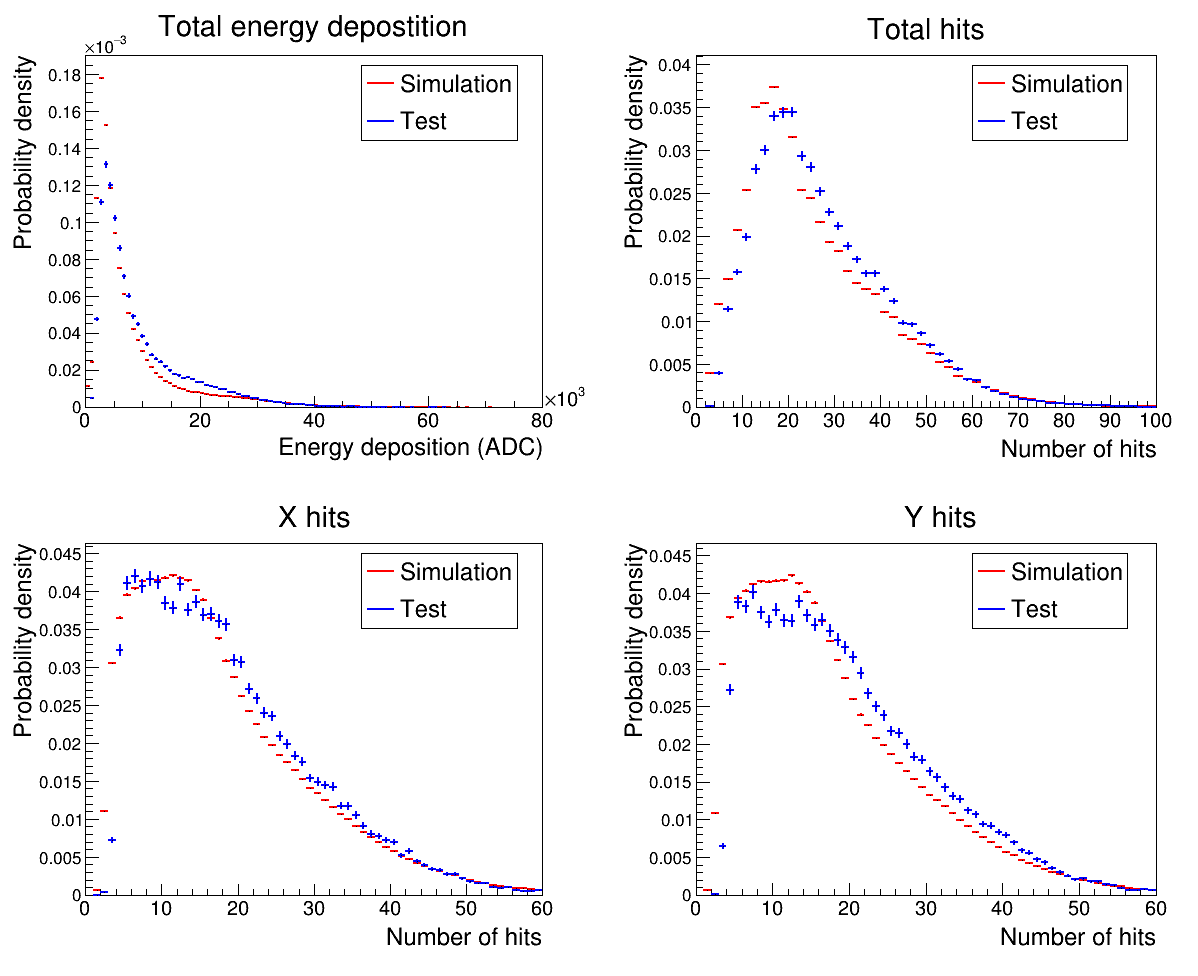}
\caption{Comparison of the distributions between simulation and test data for the $^{90}$Sr beta source. \label{fig8}}
\end{figure}

To evaluate the reliability of our simulation and digitization program, the simulation results of the $^{90}$Sr beta source data were first compared with the test data. In the simulation, the beta source was positioned \SI{5}{\centi\meter} away from the TPC entrance window, consistent with the test setup. A total of 80 million $^{90}$Sr decays, along with cascaded $^{90}$Y decays, were simulated. Since the exact amplification gain in our test setup was unknown, a gain correction factor was introduced in the simulation to calibrate the energy spectrum peak position according to the test data. Figure~\ref{fig8} shows a comparison between the energy spectrum, X-dimensional hit channels, Y-dimensional hit channels, and total hit channels from the simulation and the test data. The distributions are normalized to probability density functions for comparison. The simulation results match the test data well with only one calibration parameter. The distortion observed in the test energy spectrum, compared to the simulation, is likely due to the inhomogeneous signal response across different regions of the readout plane. This effect leads to larger fluctuations in the induced charge signal. The main cause of this inhomogeneity is the non-uniformity of the avalanche gain. A previous test of the TPC has shown that the non-uniformity is about 12.8\% within the detector central region.

\begin{figure}[htbp]
\centering
\includegraphics[width=\hsize]{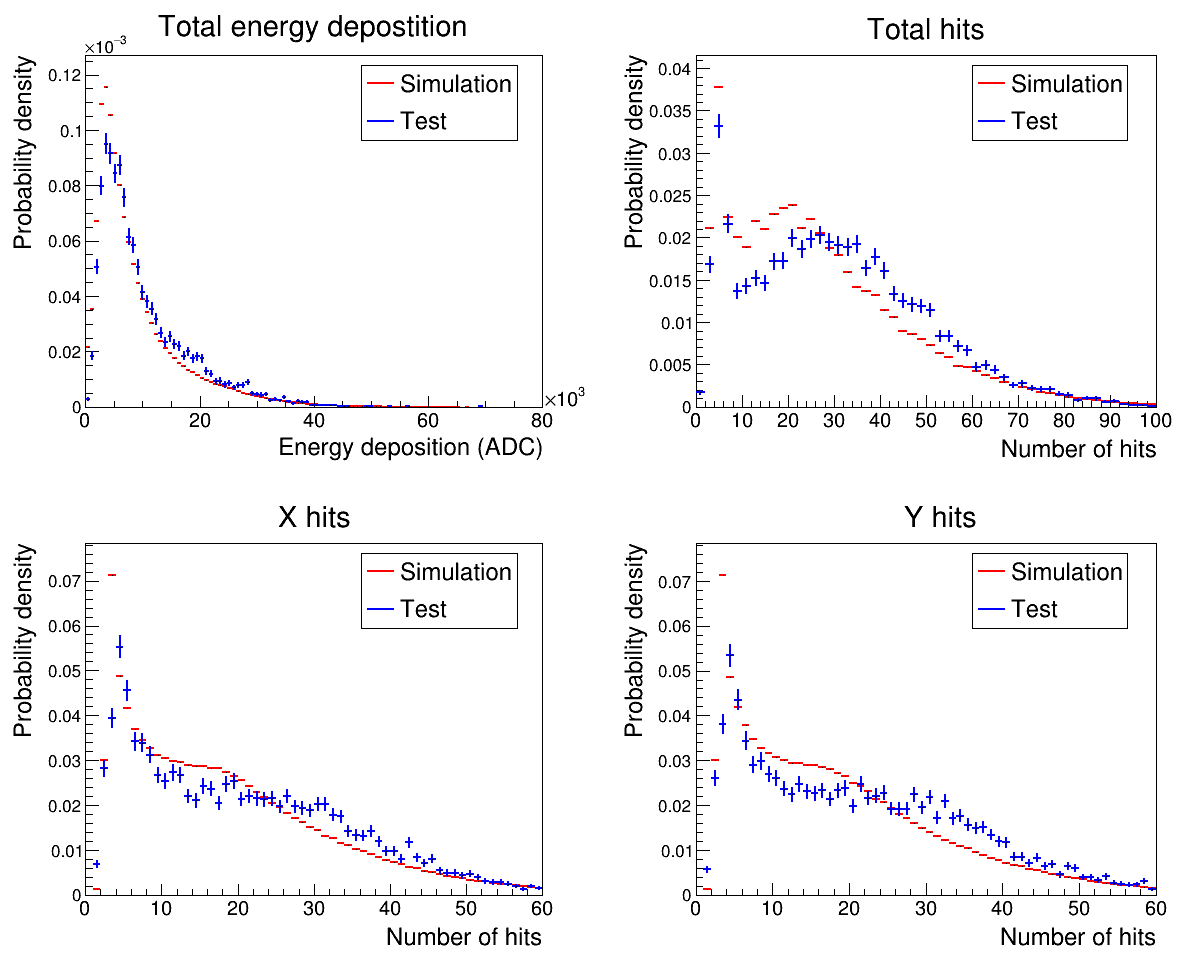}
\caption{Comparison of the distributions between simulation and test data for the background. The simulation data includes only gamma-induced background. \label{fig8-2}}
\end{figure}

In addition to the beta source events, it is also valuable to obtain information about background events through simulation. As the exact origin of the background is unknown, a reasonable assumption is that the majority of the background is caused by environmental gamma-induced electrons, given that cosmic ray muons have already been rejected by the anti-coincidence detector. Based on this assumption, we measured the gamma spectrum using a CZT detector and applied an unfolding algorithm~\cite{LIU2024169315} to obtain the true gamma flux in our laboratory. This gamma flux spectrum was then used as input for our simulation program to generate background data. Figure~\ref{fig8-2} shows a comparison between the simulated background and the test background. There are slight discrepancies in the distribution of hit channel numbers between the test results and the simulation, with the test data showing a higher frequency of events with greater hit multiplicity. This implies that additional background sources, aside from environmental gamma radiation, may be present in the experimental conditions, contributing to longer track lengths within the TPC. Nevertheless, environmental gamma-induced background is believed to be the primary source of background. Further experimental tests and analysis, as discussed in Section~\ref{bkg study}, have reinforced this conclusion and offered insights into the origins of the remaining background sources. Also, the effect of these differences has turned out to be minor on beta-background discrimination, which will be clarified in Section~\ref{sec5}.

\section{Beta and background discrimination}
\label{sec5}

Discriminating between two types of data (in our case, beta signal and background) is a classification problem~\cite{Aggarwal2015}. There are numerous algorithms developed for such problems, collectively known as multi-variable analysis (MVA) methods, many of which are implemented in the ROOT TMVA toolkit~\cite{RN41}. In this work, several TMVA methods were tested to identify the best performing algorithm. The variables used for classification include the deposited energy within the TPC, the hit strip numbers in the X and Y dimensions, the traversal time of a track inside the TPC, and the reconstructed start position on the XY plane.

\begin{figure}[htbp]
\centering
\includegraphics[width=\hsize]{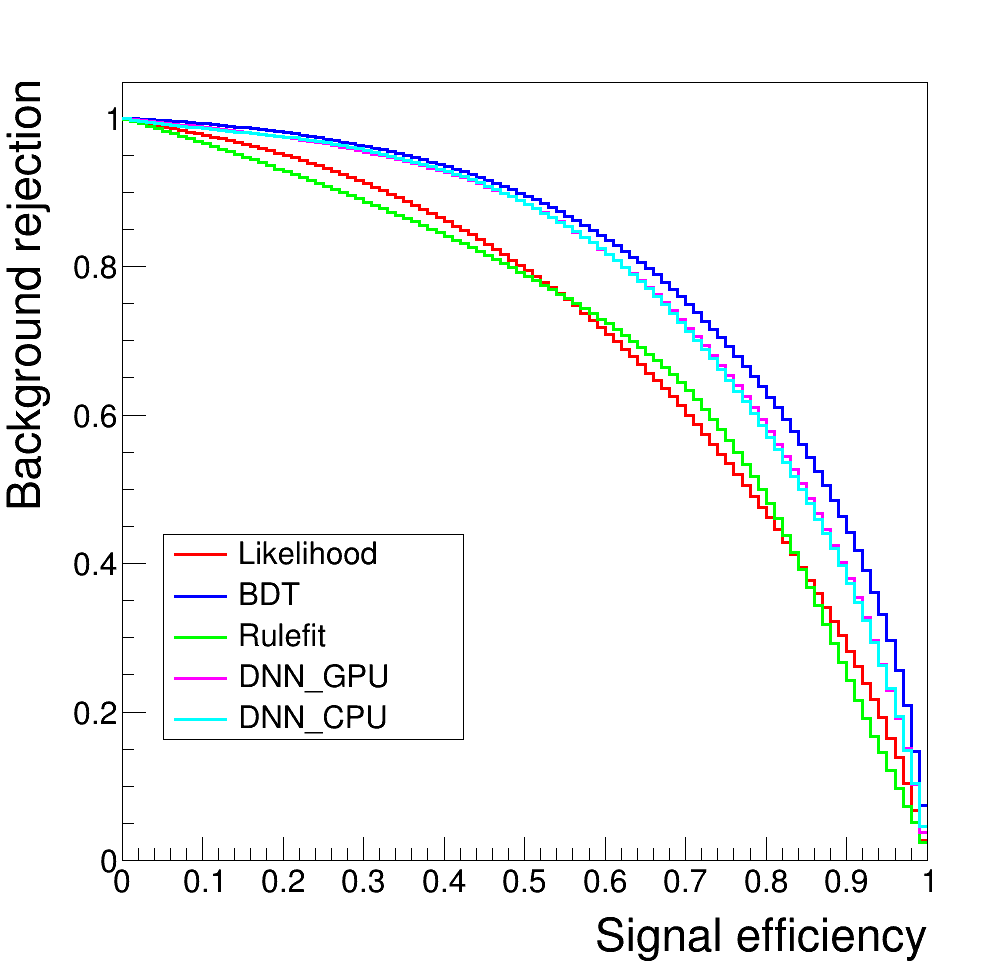}
\caption{ROC curves for different classification methods. \label{fig9}}
\end{figure}

Simulated beta and background datasets were used to test various algorithms, as we have already confirmed the consistency between the simulation and test data. A set of initial cuts was applied to both the beta signal and background events for preliminary selection. The cuts applied were as follows:
\begin{itemize}
    \item The signal amplitude must be at least ten times the noise root mean square (RMS) above the baseline. This cut helps remove small signals generated by random noise without affecting the actual data.
    \item At least one hit strip in both the X and Y dimensions is required to enable track reconstruction in both the XZ and YZ planes. The results show that nearly all events satisfy this criterion.
    \item The event should contain hits only within the TPC detector. Events with any hits on the anti-coincidence detector are considered background and are discarded.
    \item The reconstructed start position of the event should be within the central window region. This crucial cut helps reduce background from peripheral products while preserving beta detection efficiency. Only a small fraction of beta particles are rejected by this cut due to position reconstruction failure or multi-scattering.
\end{itemize}
A fiducial area with a diameter of \SI{7}{\centi\meter} was thus defined by the above cuts. These cuts retained 95.5\% of the original beta events while eliminating 81.9\% of the background events. Using the remaining data samples, several commonly used MVA algorithms were trained and evaluated. The receiver operating characteristic (ROC) curves are depicted in Figure~\ref{fig9}. The results demonstrate that the TMVA  Boost Decision Trees (BDT) method achieved the best classification performance, even without parameter tuning. This is likely because the BDT method is particularly effective at handling variables with internal correlations.
As a consequence, the TMVA BDT algorithm was selected as the most suitable discrimination method and was used in the subsequent analysis.

\begin{figure}[htbp]
\centering
\includegraphics[width=\hsize]{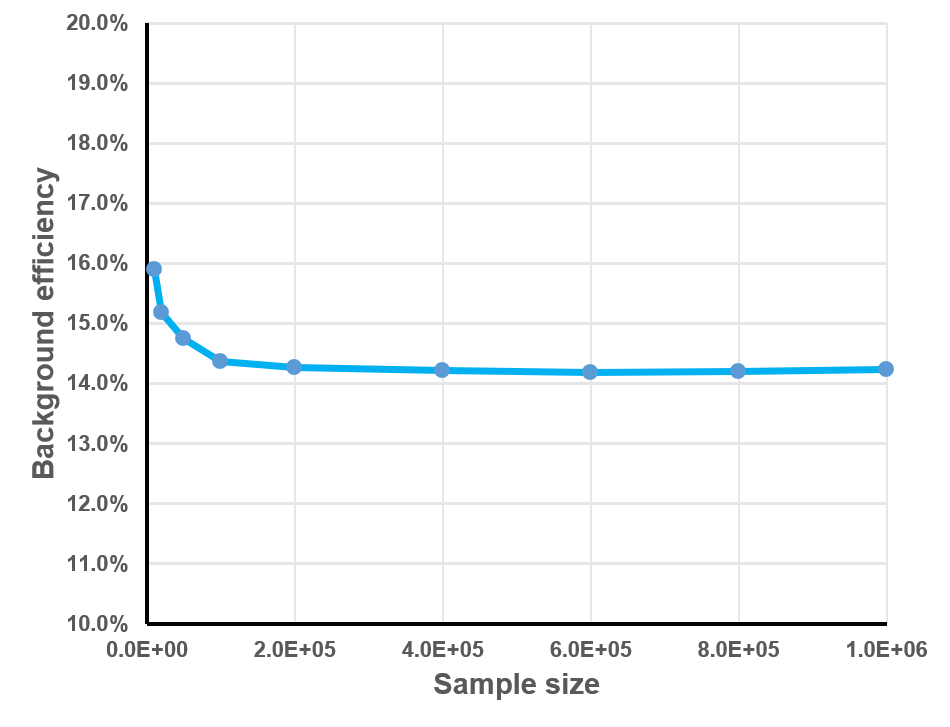}
\caption{Training performance for TMVA BDT algorithm under different sample sizes. \label{fig11}}
\end{figure}

The parameters of the BDT method were carefully tuned to maximize discrimination performance. For instance, the training sample size was varied from 100,000 to 1 million to study its impact on the classification ability, as shown in Figure~\ref{fig11}. The background efficiency at a 55\% signal efficiency was used as a metric to evaluate the discrimination quality. As the sample size increases, the discrimination performance improves initially but gradually reaches a plateau. To balance processing time and classification accuracy, a sample size of 200,000 was chosen.

\begin{figure}[htbp]
\centering
\includegraphics[width=\hsize]{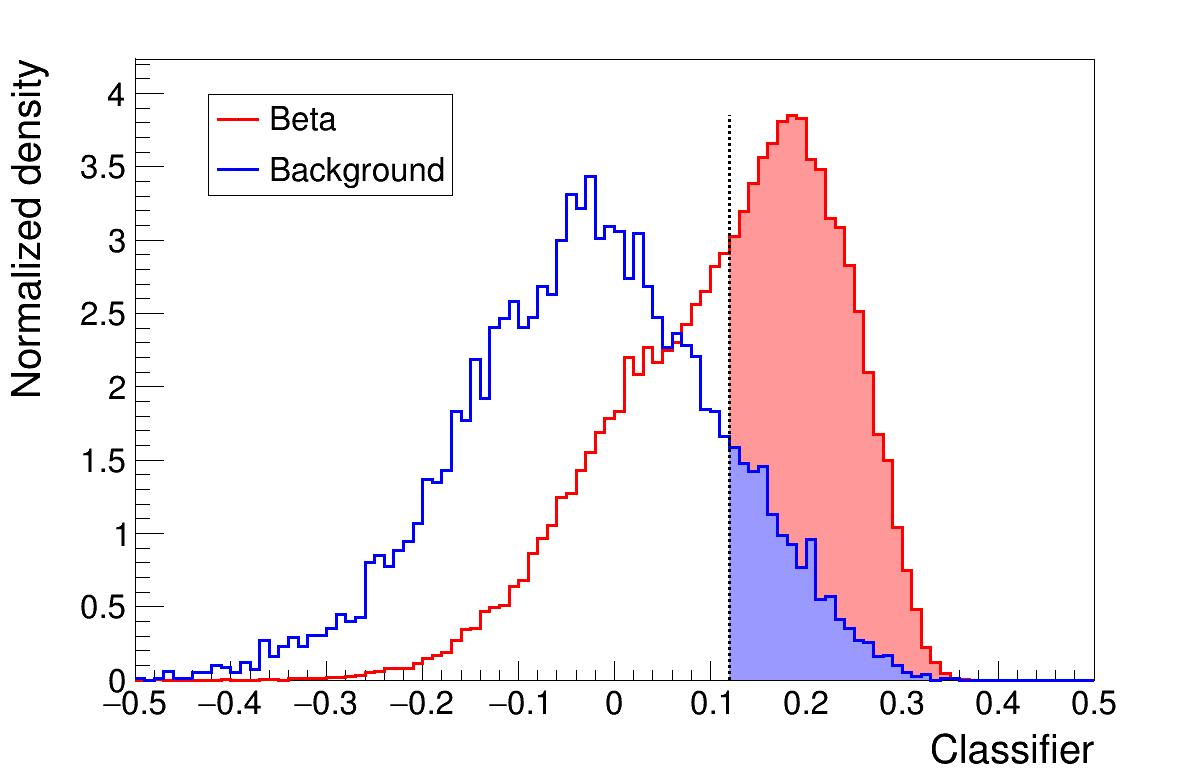}
\caption{Distribution of classifier values for the experimental beta and background samples. The dashed line marks the applied cut value. \label{fig11_2}}
\end{figure}

The trained BDT method was subsequently applied to the analysis of actual test data. A $^{90}$Sr dataset with \SI{2}{\minute} acquisition time and a background dataset with \SI{60}{\minute} acquisition time were used as signal and background samples, respectively. Since the background count rate is only about 0.6\% of the $^{90}$Sr event rate, the $^{90}$Sr dataset is considered a pure signal sample. A classifier cut value was chosen to retain a 55\% signal efficiency. Under this condition, the corresponding background rejection rate is 86.3\%, resulting in a background count rate of 18.8 counts per minute (cpm) within the sensitive area after discrimination, equivalent to \SI{0.49}{cpm/cm^2}. The resulting BDT classifier distribution, along with the applied cut value, is shown in Figure~\ref{fig11_2}. Notably, when the trained BDT model was applied to the simulated beta and background datasets, the background rejection rate was 85.7\% at the same signal efficiency. This result closely matches that obtained with the real data, indicating that the slight differences between simulation and test data are tolerable for the algorithm, and comparable discrimination performance can be achieved.


\section{Background study} \label{bkg study}
\label{sec6}

\begin{figure}[htbp]
\centering
\includegraphics[width=\hsize]{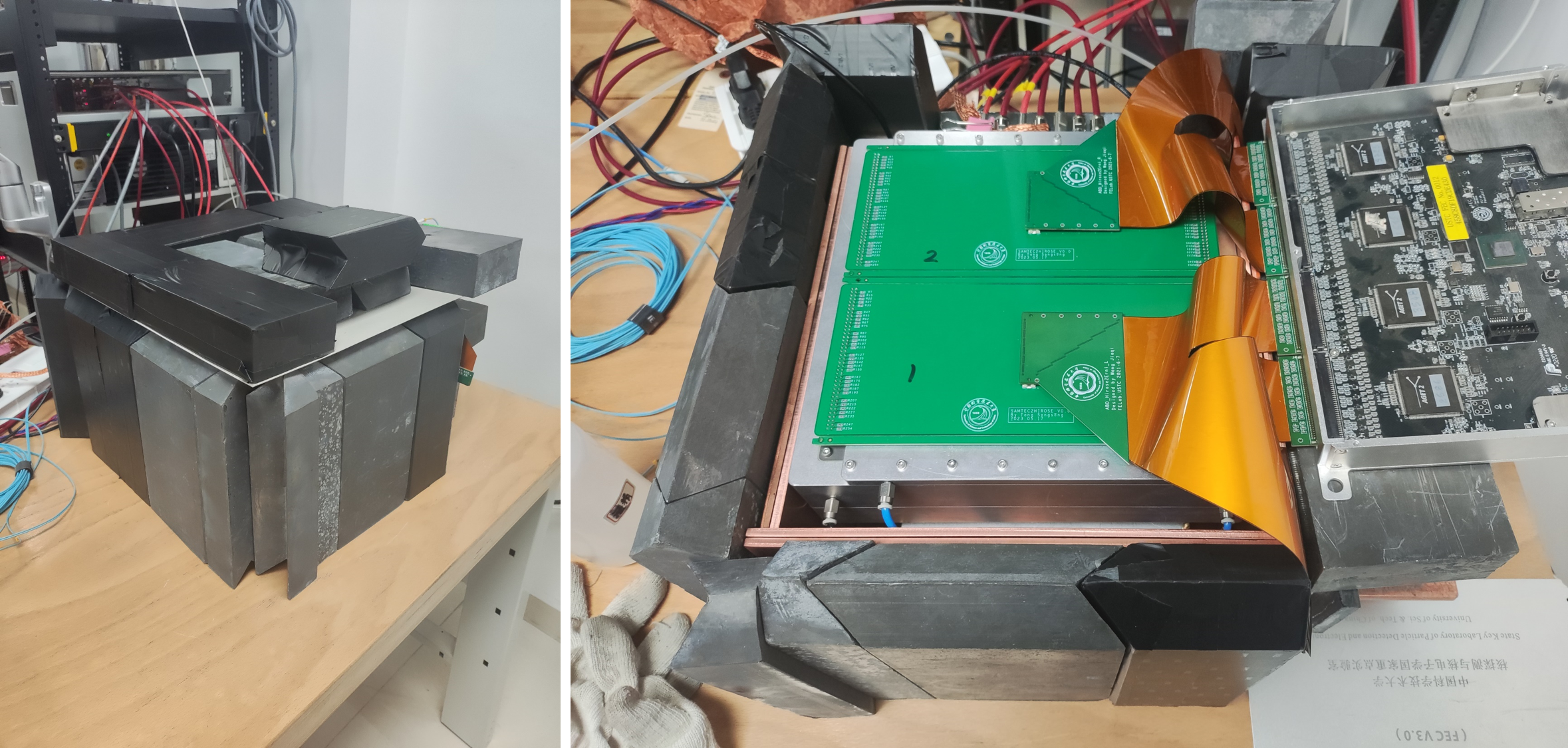}
\caption{Experimental setup for lead shielding test. \label{fig10}}
\end{figure}

To further reduce the background rate to less than \SI{0.1}{cpm/cm^2}, a detailed understanding of the origin of beta background is crucial. As mentioned in Section~\ref{sec4}, we have assumed that the primary source of background in our TPC system is environmental gamma radiation, as cosmic ray-induced penetrating background events can be rejected by the anti-coincidence detector. Our top priority is to validate this assumption. Environmental gamma radiation primarily originates from gamma-emitting isotopes present in building materials and the atmosphere, typically with energies below \SI{5}{\mega\electronvolt}. Such radiation can be effectively blocked using thick, high-\emph{Z} materials, a principle commonly applied in commercial beta detection devices to suppress background. In simulations, we observed that adding a \SI{5}{\centi\meter}-thick lead shell around the TPC system reduces the total gamma-induced background to approximately \SI{7.3e-4}{cpm/cm^2}, which is less than 0.01\% of the current value. This result demonstrated that a lead shield of this thickness is sufficient to block nearly all environmental gamma radiation. We subsequently wrapped the TPC system with \SI{5}{\centi\meter}-thick lead bricks to measure the remaining background originating from other sources (see Figure~\ref{fig10}).
Since the lead bricks used here are not radio-pure and may contain beta-emitting isotopes, an additional \SI{1}{\centi\meter} oxide-free copper board was placed between the lead and the TPC system to prevent beta particles from the lead radioisotopes from entering the TPC. The test results after shielding show that the background count rate was reduced to 30\% of its original value, reaching approximately \SI{1.04}{cpm/cm^2} (without classification). 
This finding confirms that environmental gamma radiation is the primary source of the background, which aligns with our previous background source assumption based on simulation results.

The source of the remaining 30\% of background was further investigated. Since these background events were not blocked by the lead shielding, they must originate from the materials of the TPC system itself. A plausible explanation is that beta-emitting radioactive isotopes in FR-4-based PCB materials are the primary contributors to this background~\cite{zhang2015study}. Both the anode readout board and the field cage of the TPC are made of this type of material. 

\begin{figure}[htbp]
\centering
\includegraphics[width=\hsize]{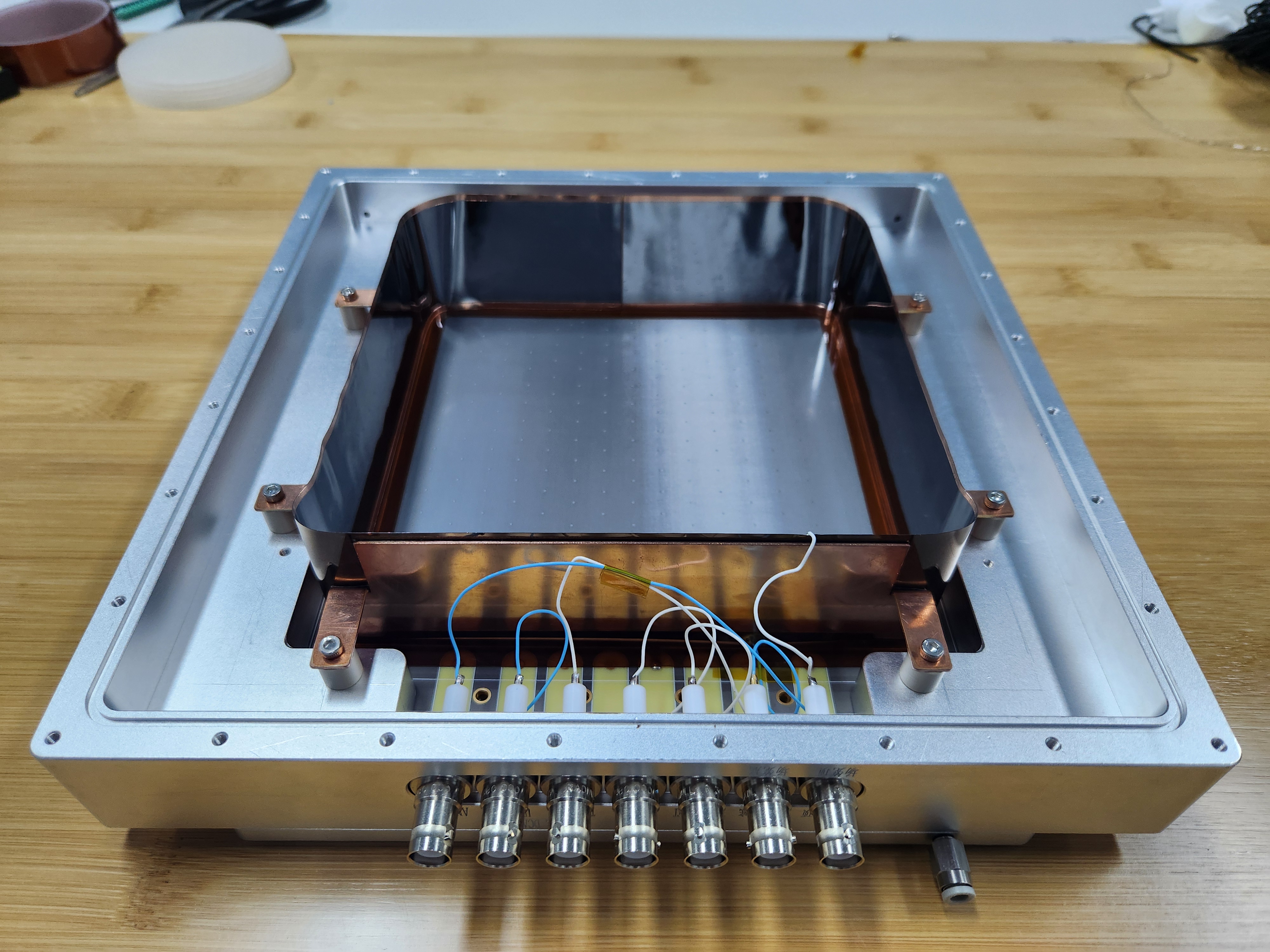}
\caption{A photogragh of the TPC with a flexible PCB field cage.\label{fig12}}
\end{figure}

To examine this hypothesis, the TPC's field cage was replaced with a low-background flexible PCB material while keeping the rest of the system unchanged for comparison (see Figure~\ref{fig12}). This optimized PCB material has been successfully used in the manufacture of low-background Micromegas detectors for the PandaX-III experiment~\cite{chen2017pandax}. \chadded[id=rzhang]{Its radioactive background was tested at the China JinPing Underground Laboratory (CJPL), demonstrating an extremely low background level of \SI{24}{\micro Bq/cm^2}~\cite{wen2024design,hpge_detector}.} Therefore, implementing this low-radioactivity field cage was expected to reduce the background contribution to a negligible level. The TPC system with the new field cage was tested under the same conditions as the original system, both with and without shielding. The background event rates for different TPC configurations and experimental setups are summarized in Table~\ref{tab2}. With the field cage material replaced by the flexible PCB, the background count rate was reduced by 15.5\% in the non-shielded setup and by 38.4\% in the shielded setup. This indicates that the original FR-4 PCB field cage contributed approximately 11.6\% to the total background in the non-shielded configuration, corresponding to about \SI{0.40}{cpm/cm^2}. 

\begin{figure}[htbp]
\centering
\includegraphics[width=\hsize]{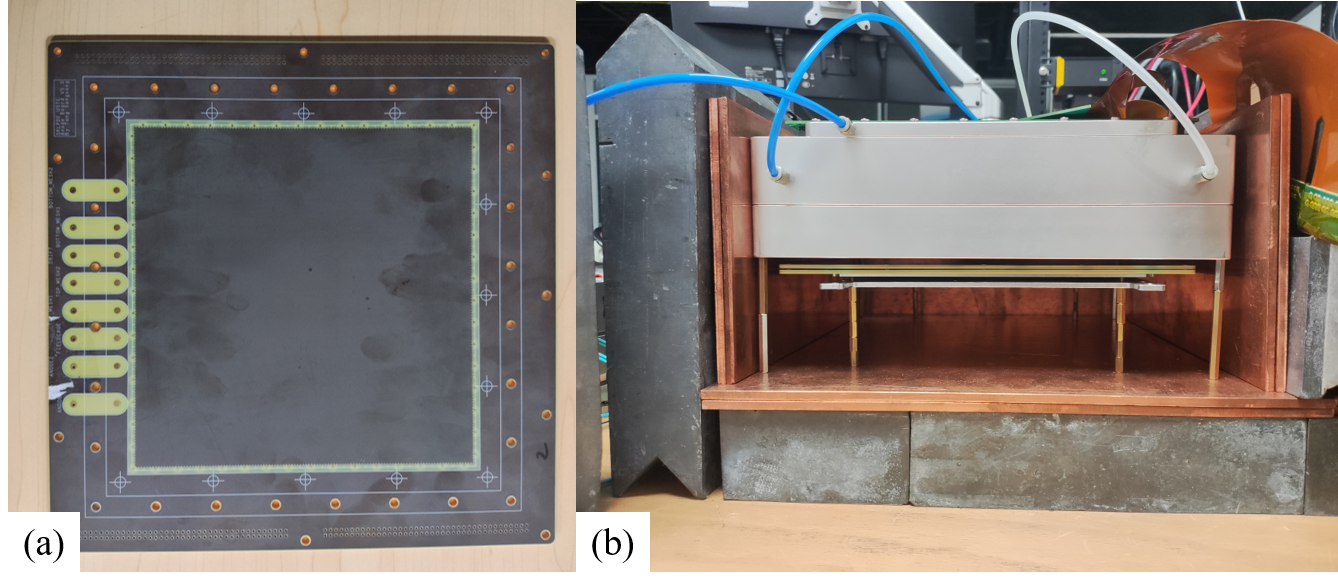}
\caption{(a). The anode PCB of the TPC system; (b). Experimental setup for the anode PCB background test, where the anode PCB is positioned on an aluminum support structure close to the detector.  \label{fig13}}
\end{figure}

To further estimate the background contribution from the anode PCB, an indirect test procedure was carried out. In this test, the TPC system was shielded, and an identical anode PCB was placed underneath the TPC entrance window (see Figure~\ref{fig13}). By comparing the background count rate in this setup to the rate without the additional PCB, the difference in background contribution was measured. An increase in the count rate is expected, as beta particles emitted from the additional anode board could penetrate the entrance window and be detected. The results shows an increase of \SI{7.4}{cpm} with the anode PCB in place. Although this measured count rate does not directly represent the actual contribution of the TPC anode PCB, due to the asymmetric layout of the two boards, it can be scaled using a conversion factor. This factor was derived by comparing simulation results for both test setups. The input for the simulation, including beta-emitting isotopes and their activities, was taken from~\cite{zhang2015study}. Using the simulated conversion factor of 3.45, the background contribution from the TPC anode PCB is estimated to be \SI{0.66}{cpm/cm^2}. 
This value closely matches the remaining background of \SI{0.65}{cpm/cm^2} after shielding and replacing the field cage material, indicating that the anode PCB is another significant contributor to the beta background. To summarize, the estimated contributions of the three primary background components, along with their respective count rates and proportions, are listed in Table~\ref{tab3}. Although other minor background sources may exist — such as radioactivity from additional detector materials or cosmic rays not vetoed by the anti-coincidence detector — their contributions to the total background are considered negligible.

\begin{table}[htbp]
\centering
\caption{Background count rates for different field cage materials with and without lead shielding.\label{tab2}}
\smallskip
\begin{tabular}{l|ccc}
\hline
Condition & \makecell{Acquisition \\ time (\si{\minute})} & \makecell{Event number \\ after cuts} & \makecell{Count rate \\ (\si{cpm/cm^2})} \\
\hline
FR-4 PCB, not shielded & 120 & 16126 & 3.49\\
FR-4 PCB, shielded & 120 & 4859 & 1.05\\
Flexible PCB, not shielded & 240 & 27259 & 2.95\\
Flexible PCB, shielded & 240 & 5982 & 0.65\\
\hline
\end{tabular}
\end{table}

\begin{table}[htbp]
\centering
\caption{Estimated background rates and their relative contributions to the total background, for the original TPC design with an FR-4 based field cage.\label{tab3}}
\smallskip
\begin{tabular}{l|cc}
\hline
Background component & Count rate (\si{cpm/cm^2}) & Fraction \\
\hline
Environmental gamma & 2.43 & 69.6\% \\
Field cage & 0.40 & 11.5\% \\
Anode board & 0.66 & 18.9\% \\
\hline
Sum & 3.49 & 100\% \\
\hline
\end{tabular}
\end{table}

\section{Detector optimization and expected performance}
\label{sec7}

\begin{figure}[htbp]
\centering
\includegraphics[width=\hsize]{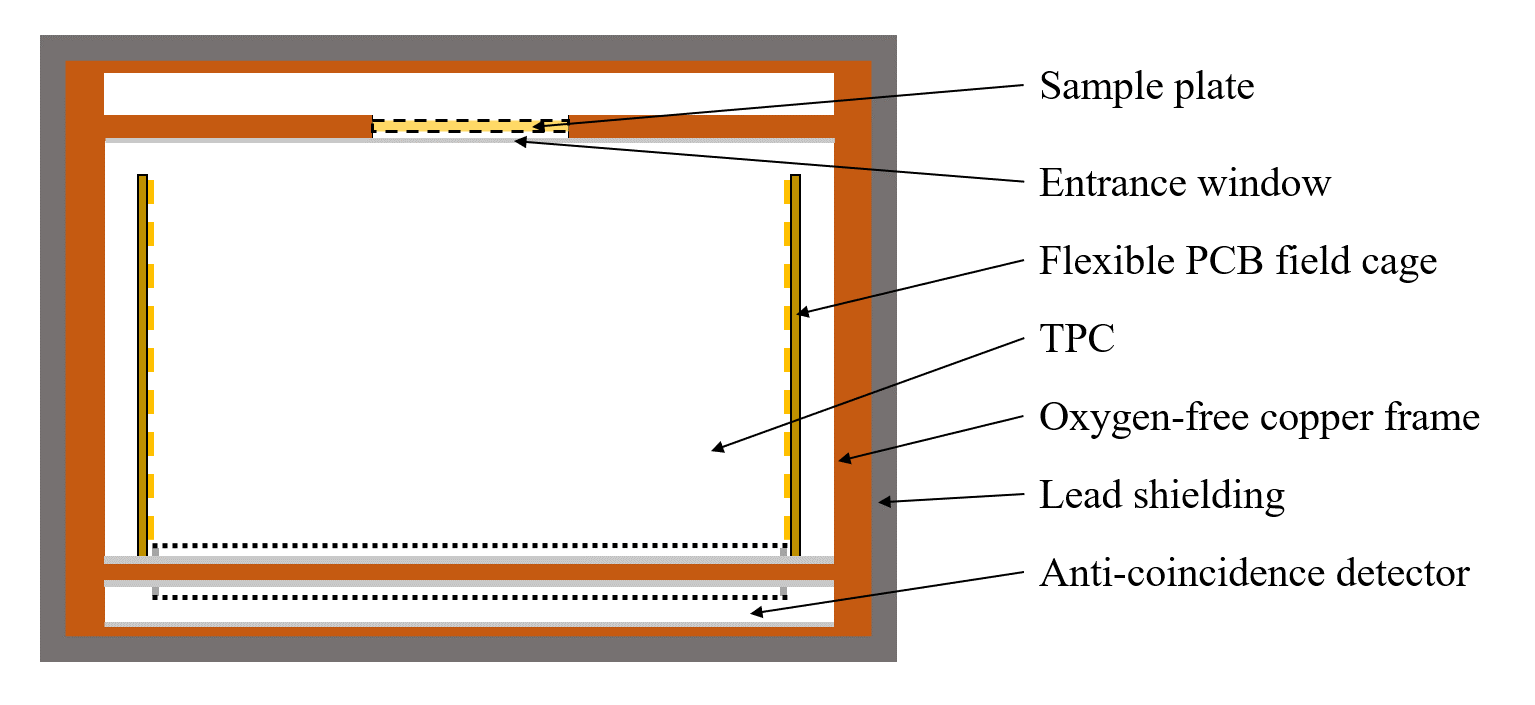}
\caption{Schematic of the optimized version of the TPC system. \label{fig14}}
\end{figure}

From the results in Table~\ref{tab3}, it is observed that environmental gamma radiation accounts for approximately 70\% of the total background, while the remaining 30\% originates from the intrinsic beta radioactivity of the PCB. It is anticipated that the PCB-induced background can be significantly reduced by replacing the FR-4 material with low-radioactivity flexible PCB material, as demonstrated by the substantial reduction in background observed when the field cage was replaced in Section~\ref{bkg study}. Therefore, the critical factor in achieving an even lower background rate lies in reducing the environmental gamma-induced background, which remains the dominant source of beta radiation. To achieve this, a new version of the TPC with an optimized detector layout was designed, aiming to strike a balance between mitigating background and maintaining detector portability. A schematic design of the optimized TPC is shown in Figure~\ref{fig14}. In addition to utilizing low-background PCB materials, the entire mechanical structure has been replaced with oxygen-free copper. This material has a higher atomic number (Z) than aluminum, and also features a lower intrinsic radiation level compared to standard copper types~\cite{copper}. To further reduce background contributions from the surrounding environment, additional shielding layers were incorporated around the TPC system. The shielding consists of an outer layer of lead with a thickness ranging from \SI{0.5}{cm} to \SI{2}{cm}, and a \SI{0.5}{cm} thick inner layer of oxygen-free copper, which serves to block any intrinsic beta decay from the lead itself.

Simulations were conducted to evaluate the expected performance of the new TPC design, focusing on the environmental gamma background, as the contribution from the flexible PCB material is eliminated. The simulation considered three different lead shielding thicknesses, with results summarized in Table~\ref{tab4}. The results indicate that with a lead shielding thickness of \SI{2}{cm}, the environmental gamma background is reduced to 0.33\% of the current level without BDT classification. With a beta signal efficiency of 55\%, the remaining background is further reduced to 0.05\%. This leads to an exceptionally low background rate of \SI{0.0012}{cpm/cm^2} in a \SI{7}{cm} diameter region.
\begin{table}[htbp]
\centering
\caption{Simulated background rates for the optimized TPC system. The last column corresponds to the value after TMVA BDT classification at 55\% signal efficiency.
\label{tab4}}
\smallskip
\begin{tabular}{l|cc}
\hline
Lead thickness & \makecell{Count rate\\ before classification\\ (\si{cpm/cm^2})} & \makecell{Count rate\\ after classification\\ (\si{cpm/cm^2})}\\
\hline
\SI{0.5}{cm} & 0.090 & 0.015 \\
\SI{1}{cm} & 0.035 & 0.0058 \\
\SI{2}{cm} & 0.008 & 0.0012 \\
\hline
\end{tabular}
\end{table}

\section{Conclusion}

In this paper, a TPC prototype for beta radiation detection has been designed and fabricated. The TPC is equipped with a Micromegas detector to eliminate cosmic ray background through anti-coincidence. The test results demonstrated that the TPC system is capable of reconstructing beta particle tracks and energy deposition. A simulation program was developed to assess the performance of this TPC system, including a digitization process to convert the simulation results into signal waveforms. The simulated data aligns well with the test data, according to the distribution of various variables. Several multi-variable analysis models were evaluated to discriminate between beta and background data, with the TMVA BDT model selected for its superior performance. The BDT method, trained using simulation datasets, was then applied to classify test data. With this discrimination method, the TPC system achieved a background rate of \SI{0.49}{cpm/cm^2} without shielding, while retaining 55\% of the beta events. The source of the beta background was investigated by comparing the results under various detector settings, revealing that environmental gamma radiation, anode board radioactivity, and field cage radioactivity are the three primary contributors to the beta background. 

An optimized version of the detector, featuring low-background flexible PCB material and oxygen-free copper, has been designed to further minimize the background. Simulation results indicated that this new TPC design is expected to achieve a total background rate of \SI{0.0012}{cpm/cm^2} with a beta signal efficiency of 55\%. The new version of the TPC holds promise for practical applications.


\section{Acknowledgments}

This work was partially performed at the University of Science and Technology of China (USTC) Center for Micro and Nanoscale Research and Fabrication, and we thank Yu Wei for his help in the nanofabrication steps of the germanium coating for the resistive anode. For the development of the TPC prototype, we are grateful to Dr. Sicheng Wen, Dr. Anshun Zhou, and the R\&D team of JIANWEI Scientific Instruments Technology Co., Ltd.


\end{document}